\documentclass[aps,floatfix,showpacs,amsmath,superscriptaddress,nofootinbib,10pt,a4paper]{revtex4-1}
\usepackage{graphicx}
\newcommand{\be}{\begin{equation}}
\newcommand{\ee}{\end{equation}}
\newcommand{\bea}{\begin{eqnarray}}
\newcommand{\eea}{\end{eqnarray}}

\begin{document}

\title{Merger Dynamics in Three-Agent Games}

\author{Tongu\c{c} Rador}\email[]{tonguc.rador@boun.edu.tr} 
\affiliation{Bo\~{g}azi\c{c}i University, Department of Physics \\ Bebek 34342, \.Istanbul, Turkey}
\affiliation{\.Izmir Institute of Technology, Deparment of Physics \\ Urla 35430, \.Izmir, Turkey}
\author{R\"{u}\c{s}t\"{u} Derici}\email[]{rustu.derici@isbank.com.tr}
\affiliation{\.I\c{s} Bankas\i, \.{I}\c{s} Kuleler, Kule 1 \\ Levent 34330, \.Istanbul, Turkey}

\begin{abstract}
We present the effect of mergers in the dynamics of  the three-agent model studied by Ben-Naim, Kahng and Kim  and by Rador and Mungan. Mergers are possible in  three-agent games because two agents can combine forces against the third player and thus increase their probability to win a competition. We implement mergers in this three-agent model via resolving merger and no-merger units of competition in terms of a two-agent unit. This way one needs only a single parameter
which we have called the competitiveness parameter. We have presented an analytical solution in the fully competitive limit. In this limit    the score distribution of agents is stratified and self-similar.  
\end{abstract}


\maketitle

\section{Introduction}

The use of methods inspired from physical principles has recently been of wide utility in studying various systems. Models with  explanatory  and predictive powers have been applied to biological, social, political, economical, animal behavioral and many other phenomena. The common
property of all these models is that they all involve a collection of agents interacting with a well defined set of rules. The set of
rules may describe a competition for earning a game, for exchanging an attribute say money, energy or points, or to fight for a position in space. 

Recently such a model based on two-agent units of competition is introduced and was applied to sports data\cite{onlar1}-\cite{onlar3}. Later  that model is generalized to three-agent units providing qualitative understanding of emerging social structures \cite{onlar4}. That model being deterministic
did cover the full range of possibilities. This extension is achieved in  \cite{biz} where the full phase space of three-agent games is found yielding new social structures.

In the present paper we study an intriguing extension of the three-agent model ; the possibility that two agents could combine forces against the third one, in other words merger dynamics. In \cite{onlar4} and \cite{biz} it was shown that the competitive subspace of three-agent games yields a continuous point gain rate for agents. Implementing mergers into the model we find that the  agents condense at particular
values of income rate. That is, as a result of mergers the society of agents becomes {\bf stratified}. In the limit where the model becomes
the most competitive the distribution for income rate becomes self-similar.

We first review the main results of \cite{biz} to introduce the three-agent model. The remaining parts of the manuscript is devoted to the study of merger dynamics.

\section{Review of Three Agent Games}
In this chapter we will review the dynamics of three agent games to have a structured manuscript. We shall omit various
details as those were already discussed in depth in \cite{biz}. We shall however add some aspects that are not discussed in the
mentioned paper and we shall add emphasis on points relevant for the next chapter where we will implement mergers into the model.

Let us consider a collection of $N$ agents with a given distribution of points. We pick three of them randomly and order their
points say as $L\geq M\geq S$. We shall give one point to one and only one agent according to the following rules,

\begin{list}{$\bullet$}{}
\item{} $\left\{L>M>S\right\}$    $\Longrightarrow$      $\left\{P, \;T, \;Q\right\}$
\end{list}
\begin{list}{$\circ$}{}
\item{} $\left\{L=M>S\right\}$    $\Longrightarrow$      $\left\{\frac{P+T}{2},\; \frac{P+T}{2},\; Q\right\}$
\item{} $\left\{L>M=S\right\}$    $\Longrightarrow$      $\left\{P,\; \frac{T+Q}{2},\; \frac{T+Q}{2}\right\}$
\item{} $\left\{L=M=S\right\}$    $\Longrightarrow$      $\left\{\frac{1}{3},\; \frac{1}{3},\; \frac{1}{3}\right\}$
\end{list}

\noindent Here the lists on the right represent the winning probabilities of the teams with points listed on the left. In view of later
chapters we refer to this model as the {\em single rule model} emphasizing the fact that there are no conditions on the points other than their ordering. As can be inspected, cases when some agents have equal points are resolved on the basis of {\em equal likelyhood}. That after every game one agent surely advances requires the normalization of the probabilities

\be
P+T+Q=1\;.\nonumber
\ee 

Now let us denote the probability to pick an agent with point $x$ at a particular time as $f_{x}$, this should really be taken as a shorthand notation for $f(x,t)$.  After a competition some teams might leave this region towards $x+1$ and some teams might enter it from $x-1$ by winning a competition in either case. This suggests the following local conservation law

\be{\label{eq:1}}
\frac{df_{x}}{d t}=\sum_{y,y'}f_{x-1}f_{y}f_{y'}\;W(x-1,y,y')-\sum_{y,y'}f_{x}f_{y}f_{y'}\;W(x,y,y')\;\;.
\ee

\noindent Here $W(x,y,y')$ denotes the probability that the agent with point $x$  will win againts two others with points $y$ and $y'$.  The microscopic rules above completely define what $W$ is.  Furthermore, the right hand side is cubic in $f$. This is so  because we are picking three agents out of the collection and the probability to pick a team with a given point $x$ is $f_{x}$. 

Since 

\be
\sum_{x}f_{x}=1, \nonumber
\ee

\noindent it is immediate that (\ref{eq:1}) also implies, as it should, the global conservation of the total number of teams, as can be checked by performing a sum over 
$x$. 

The time variable  in (\ref{eq:1}) has an arbitrary scaling which can be compensated by an overall factor in the definition of $W$ since the former represents in essence a rate. The natural scale is such that the average points of teams is given by,

\be
\bar{x}(t)\equiv\sum_{x=0}^{\infty}x\;f_{x}=\frac{t}{3}\;,
\label{eqn:xave1}
\ee

\noindent meaning that (on average) each team participates in a single game while we increment the time variable by one unit. As only one of the participating teams in a game wins and advances its score by one, equation (\ref{eqn:xave1}) follows. One can equivalently say that
the average speed with which agents increase their points is $1/3$. This normalization also means that the maximun theoretical point an agent could have acquired at time $t$ is simply $t$.

The presence of sums over the discrete indices on the right hand side of (\ref{eq:1}), 
results in a coupled set of differential equations. For the model at hand these can be much simplified by defining

\be
F_{x}\equiv\sum_{y=x^{*}}^{x-1}f_{x}\;\;.
\ee

\noindent Here $x^{*}$ represents the smallest point below which there are no agents. Since agents do not loose points this
value does not change in time and is defined by the initial point distribution as $f_{x}=0$ at $t=0$ for $x\leq x^{*}$. Without loss of generality on can take $x^{*}=0$ and confine the points to positive values. Also from the
definition of $F_{x}$ the following is immediate

\be
f_{x}=F_{x+1}-F_{x}\;.
\ee

Summing (\ref{eq:1}) over $x$ we obtain

\be\label{eq:2}
\frac{dF_{x}}{dt}=-f_{x-1}\sum_{y,y'}W(x-1,y,y')\;f_{y}\;f_{y'}\;.
\ee

We note that 

\be\label{eq:sum}
f_{x-1}\sum_{y,x}W(x-1,y,y')\;f_{y}\;f_{y'}
\ee 

\noindent yields the probability that a team with score $x-1$ will win any possible choice of single competition with two other teams. Working out the sum using the rules we get the final form of the master equation,

\bea\label{eq:harbi}
\frac{dF_{x}}{dt}=&&-f_{x-1}\left[ pF_{x-1}^{2}+q(1-F_{x})^{2}+2tF_{x-1}(1-F_{x})  \right] \nonumber \\
&&-2\frac{(p+t)}{2}f_{x-1}^{2}F_{x-1}\nonumber \\
&&-2\frac{(t+q)}{2}f_{x-1}^{2}(1-F_{x})\nonumber\\
&&-\frac{1}{3}f_{x-1}^3\;.
\eea

\noindent The terms on the first line of (\ref{eq:harbi}) represent the bulk of interactions between three players with different scores. The second and third lines represent the cases where two players have identical scores $x$ and the last term represents the case when all the teams have equal score. The effect of terms denoting units of competition where one picks some agents with equal score are irrelevant for the late time dynamics of the system as they die out in time. We will call these type of terms the {\em interface} terms and the rest as the {\em bulk} terms. So as time goes by, in a thermodynamic limit where the number of teams ranges to infinity, the majority of the contributions to the dynamics will be governed by the bulk terms. On the other hand as time goes by, almost every team will accumulate a certain number of points which, in general, will be larger than a single point. These considerations allow one to go to a continuum limit where the terms like $F_{x-1}$ are expanded in terms of the derivatives. A first order approximation, where one considers only the bulk terms, results in the following,

\be{\label{eq:cont1}}
\frac{\partial F}{\partial \tau}=-\frac{\partial F}{\partial x}\; G'(F)=-\frac{\partial G(F)}{\partial x}\;,
\ee

\noindent with 

\be{\label{eq:cont2}}
G'(F)\equiv PF^{2}+2TF(1-F)+Q(1-F)^2\;.
\ee

\noindent We will refer to this approximation as the {\em hydrodynamical limit}. We would like to mention that with this limit we still have $\bar{x}=t/3$, which corraborates the neglection of the interface terms. 

At this point we would like emphasize that the hydrodynamical
limit for {\bf any model} involving the competition of three agents will take the form

\be
\frac{\partial F}{\partial t}=-\frac{\partial F}{\partial x}\int\int dy\;dy'\;\frac{\partial F}{\partial y}\frac{\partial F}{\partial y'}W(x,y,y')\;,\nonumber
\ee

\noindent for a given function $W(x,y,y')$.

In \cite{biz} the solutions to (\ref{eq:cont1}) was studied via the method of characteristics, for all cases satisfying $P+T+Q=1$, with the initial condition $F(x,0)=\Theta(x)$ meaning that all agents starts with zero points. As is well known this is the Riemann problem.  
The asymptotic solutions was found to have the form $F(x,t)=F(x/t)$ as it is usually the case with hyperbolic equations coming from conservation laws. In
our normalization of the time variable the maximum possible theoretical point an agent can have at time $t$ is simply $t$, thus $z\equiv x/t$ satisfies $0\leq z\leq 1$. Furthermore one can show that $\bar{x}=t/3$ yields

\be
\int_{0}^{1}dz F(z)=2/3\;.\nonumber
\ee

In terms of the variable $z=x/t$ one can recast (\ref{eq:cont1}) into

\be
\frac{dF}{dz}\left[-z+PF^{2}+2TF(1-F)+Q(1-F)^2\right]=0\;\;,
\ee

\noindent subject to the requirements, which we have discussed, that $F$ has to satisfy.  The solutions fall into the following categories

\begin{list}{{\bf Regime}}{}
\item $C^{-}\,$: for  $P>T \ge Q$ and $T < 1/3$, 
\item  $C^{0}\;\,$: for $T = 1/3 > q$, 
\item $C^{+}\,$: for  $T > 1/3$ and $P \ge T$,
\item $C^{+}_{S}\,$: for  $P< T$ and $Q \le 1/3$, 
\item $S\;\;\;\,$: for   $Q > 1/3 > P$, 
\item $C^{-}_{S}\,$: for  $Q > T $ and $P > 1/3 $.
\end{list} 

\noindent Here the superscripts $+,-$ refer to the fact that $F(z)$ has a positive or negative second derivative respectively. The superscript $0$ refers to a linear curve\footnote{This case is identical to a two-agent model. Since the middle agent advances with the mean speed of the whole collection of agents, it is as if it is not there.}. The subscripts $S$ mean that the solution has a shock singularity. This happens when $F(z)$ becomes double valued in $z$ and hence the solution must be resolved via a jump much like the Maxwell reconstruction in thermodynamics. If $F$ starts at $F_{l}$ to the left of the shock and ends at $F_{r}$ to the right of it the location of the jump is given by the Rankine-Hugoniot condition

\begin{equation}
 z_{S} = \frac{G(F_l) - G(F_r)}{F_l - F_r } = \frac{\Delta G}{\Delta F}\;,
\end{equation}

\noindent which is simply restating the fact that the number of agents leaving the shock is equal to those that enter it. In geometrical
terms this means that the area between the curve $F(z)$ and $z=z_{S}$ to the left of the shock is equal to that area to the right of
the shock.

\begin{figure}[t]
\includegraphics[scale=0.5]{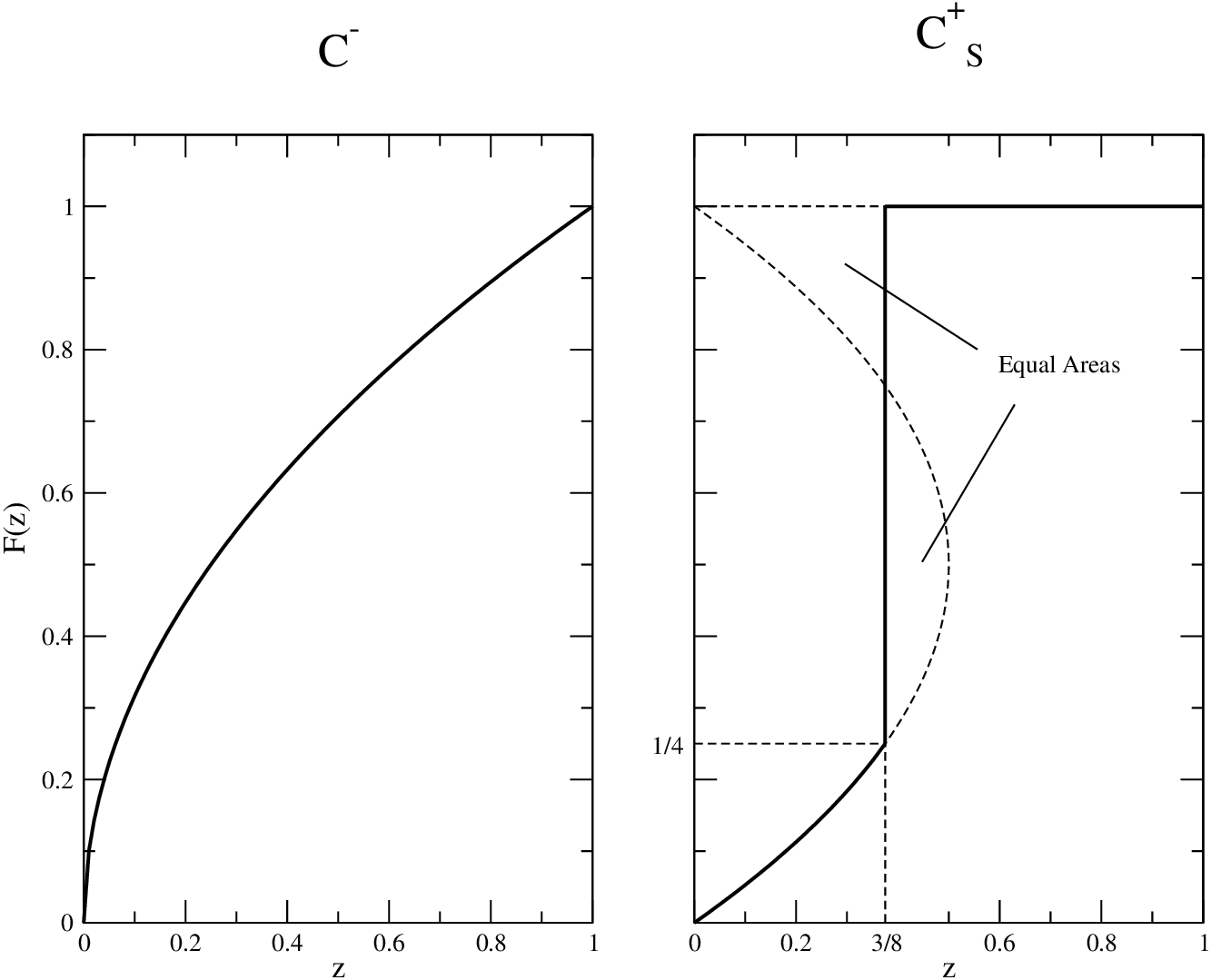}
\caption[]{Representative solutions of the single rule model relevant to our discussion. For $P=1$ we have an example of $C^{-}$ solution which in this instance is simply $F(z)=\sqrt{z}$. For $T=1$ we have an example of $C^{+}_{S}$ regime; the curve is $F(z)=(1-\sqrt{1-2z})/2$  upto the value of $z_{S}=3/8$ given by Rankine-Hugoniot condition. There is a single jump here, so we have $F_{l}=1/4$ and $F_{r}=1$. The effect of the jump condition can be interpreted geometrically as shown in the plot to the right. }
\label{fig:single1}
\end{figure}

We shall not study all the various types of solutions mentioned above. For our discussion on mergers we shall only need $C^{-}$ and $C^{+}_{S}$. We refer the reader to Fig.\ref{fig:single1} where $C^{-}$ is given for $P=1$ and $C^{+}_{S}$ is given for $T=1$.
 
\section{Merger Dynamics}

The model presented in the previous section has a single set of rules (the set $\Sigma\equiv\{P,T,Q\}$) that only depends on the ordering of  points. 
A straightforward generalization of this model can be to extend the rules so that we have various sets of probabilities
and to provide a selection rule  so as to define which set should be applied for a particular choice of agents; this selection rule
should of course be a condition on the points of the agents other than their ordering. As we shall discuss below allowing mergers falls in a small subset of such generalizations. 

\subsection{Implementing Mergers}

The idea of mergers is only meaningful if the game is competitive, where higher points are favored. 
In this respect two agents can merge points, act as a single agent against the third one. 
If the mergees loose, the third player gets the point. If on the other hand the mergees eliminate the third player we still have to 
resolve a single winner. These considerations already hint at the necessity to construct
the rules of the three-agent step in terms of a two-agent unit of competition. If this is so achieved, after 
winning against the third player the mergees can  turn against each other and play the same two-agent unit to 
decide which one of them will receive the point. 

However, if the selection rule does not  
allow a merger, we have a generic three-agent step which we should take to be competitive as well 
to be in accord with the philosophy of the idea of mergers. Such a set of rules too can be established in terms of a 
two-agent unit of competition. In \cite{biz} it was shown that a natural way to achieve this
is to let each three agents play a single two-agent match with each other: that is to have a tournament. 
All the two-agent matches  in this tournament are decided based only on how many whole tournaments the agents have won before: 
 during the tournament the tournament wins of each team (which is simply the associated 
points $L$, $M$ and $S$ of the participants) is kept constant but they accumulate match points depending on the tournament wins. 
The winner of the tournament is the agent with largest number of accumulated match points. As usual ties are decided on the basis of  equal likelihood.

\begin{table}[t]
\begin{tabular}{|c|c|c|c|c|c|c|c|}
\hline
& NO-MERGER &LM MERGER & LS MERGER& MS MERGER& MS MERGER & MS MERGER\\
 &  & $L+M>S$ & $L+S>M$ & $M+S<L$& $M+S>L$ &$M+S=L$\\\hline
$P$ & $p^2+pq/3$ & $p^2$ & $p^2$ & $p$ & $q$ &$1/2$\\\hline
$T$ & $4pq/3$ & $pq$ & $q$ & $pq$ & $p^2$ &$p/2$\\\hline
$Q$ & $q^2+pq/3$ & $q$ & $pq$ & $q^2$ & $pq$ &$q/2$\\\hline
\end{tabular}
\caption{\label{table:tab1} Winning probabilities of the three agents with points $L>M>S$ for no-merger tournaments and for various possible merger cases in terms of the competitiveness parameter $p$. Cases where there are equalities of points can be read from this table using the
equal likelyhood prescription described in Chapter II. For instance if $M=S$ the probability to win for each is $(T+Q)/2$.}
\end{table}

So we have established how to implement mergers and no-mergers for each particular three-agent competition in terms of the rules of a two-agent unit. Having approached
the problem in this way we need to use only a single parameter; the winning probability of the higher score agent in the two-agent unit competition. Let the rules for the two-agent unit  be $\{l>s\}\rightarrow \{p,q\}$: the agent with point $l$
shall get the match point with probability $p=1-q$. This unit  is competitive if $p>1/2$. After straightforward analysis we arrive at Table.\ref{table:tab1}.

We see that various cases of mergers are possible and we need a selection rule as we have argued. The simplest and most natural selection mechanism which is also
in accord with the overall competitive nature of our model is to {\bf let two agents merge if and only if both their probabilities to win the unit of competition increase
with respect to the no-merger case}. We assume that our agents are not smart and incapable of a long term strategy; they simply respond to 
an increase in the probability to win the three agent unit at hand. From the table it is evident that the agent with the highest point $L$ will never find it
profitable to merge neither with $M$ nor with $S$. Also $S$ will find it not feasible to merge with $M$ if $M+S<L$. This leaves us with MS mergers with either $M+S>L$ or $M+S=L$.
Using the selection rule stated   we find
that if $p>3/4$ mergers are feasible for those agents participating in it\footnote{In fact $M+S>L$ case only requires $p>3/5$. The higher value $3/4$ quoted in the text comes from the $M+S=L$ case. 
This term is an interface contribution and thus will represent terms with higher derivatives or higher powers of the first derivative of $F$. Consequently this term is less and less important in the asymptotic future of the sytem. However in a simulation it occurs in the early stages
if all the agents starts out with the same points so it is honest to include it in the analysis of the selection rule.}. Analyzing further we can also discover that a merger unit is rejected if $L=M>S$, we remind the reader however that  this is an interface term.

With the analysis above we have full knowledge of the function $W(x,y,y')$ introduced in the previous chapter. Other than this difference the mathematical setup of the merger model is completely analogous to the single rule model of the previous chapter. 
\subsection{The Model}

The analysis of the previous section amounts to the following rules, where to be explicit at the expense of being redundant, all the details (including the surface terms) of the model are summarized.

\begin{list}{$\bullet$}{}
\item{$L>M>S$ and $M+S<L$ no merger unit with rules $\Sigma=\left\{P,T,Q\right\}$.}
\item{$L>M>S$ and $M+S>L$ merger unit with rules $\Sigma'=\left\{P',T',Q'\right\}$.}
\end{list}
\begin{list}{$\circ$}{}
\item{$L>M>S$ and $M+S=L$ merger unit with rules $\Sigma''=\left\{P'',T'',Q''\right\}$.}
\item{$L=M>S$ no merger unit with rules     $\left\{\frac{P+T}{2},\frac{P+T}{2},Q\right\}$.}
\item{$L>M=S$ and $M+S<L$ no merger unit with rules $\left\{P,\frac{T+Q}{2},\frac{T+Q}{2}\right\}$. }
\item{$L>M=S$ and $M+S>L$ merger unit with rules $\left\{P',\frac{T'+Q'}{2},\frac{T'+Q'}{2}\right\}$.}
\item{$L>M=S$ and $M+S=L$ merger unit with rules $\left\{P'',\frac{T''+Q''}{2},\frac{T''+Q''}{2}\right\}$.}
\item{$L=M=S$ equally likely unit with rules $\left\{ \frac{1}{3},\frac{1}{3},\frac{1}{3}\right\}$.}
\end{list}

\noindent with the definitions $P=p^{2}+pq/3$, $T=4pq/3$ and $Q=q^{2}+pq/3$ for no-merger units, $P'=q$, $T'=p^{2}$ and $Q'=pq$ for merger units
with $M+S>L$ and $P''=1/2$, $T''=p/2$ and $Q''=q/2$ for merger units with $M+S=L$. 

It is important to note that if only the rules $\Sigma$ or $\Sigma'$ are applied unconditionally during the simulation, the resulting distribution is in the $C^{-}$ or $C^{+}_{S}$ regime respectively of the single rule model. 
As the reader could have already guessed, this is
somewhat evident since the set $\Sigma$ favors the leading agent whereas the set $\Sigma'$ is preferring the middle agent. Therefor
when mergers are implemented the two rules are in conflict with each other. This effect is much more pronounced 
if we let the system approach the limit of full competitiveness by
letting $p\to 1$ in which case probabilities converge to $\Sigma=\{1,0,0\}$ and $\Sigma'=\{0,1,0\}$ \footnote{One may object to this limit by observing that
the probability to win for the agent with the lowest point remains zero and thus this player has no incentive to participate in a merger with the middle agent. However
letting $p=1-\theta$ and expanding the probabilities to first order we see that $Q=\theta/3$ and $Q'=\theta$. Thus the lowest lagger triples its chances
to win in merging no matter how close to zero its winning chances are. }. 
We thus expect on general grounds that, if there is no prevalence of a single rule in the game 
the resulting dynamics of the system
should emerge from this conflict as something that is  neither a $C^{-}$ nor a $C^{+}_{S}$ solution but a new state which bears aspects of both.

Now we have, as usual,

\be
\frac{\partial F_{x}}{\partial t}=-f_{x-1}\sum_{y}\sum_{y'}W(x-1,y,y')f_{y}f_{z}\; .
\ee

\noindent However in contrast to the single rule model the sum above is very complicated which we do not duplicate here. In the hydrodynamical limit we still have,
\be
\frac{\partial F}{\partial t}=-{\mathcal G}'[F]\frac{\partial F}{\partial x}\;.
\ee
\noindent The particulars of it, however, are complex as expected (we have suppressed the time dependece of $F$'s to have a readable expression):

\bea\label{eq:yuh}
\mathcal{G}'[F]&=&P'\left[F^{2}(x)-F^{2}(x/2)\right]+PF^{2}(x/2)+2(P-P')\int_{x/2}^{x}dy\frac{\partial F}{\partial y} F(x-y)\nonumber \\
	      &+&2T'\left[F(2x)-F(x)\right]F(x)+2T\left[1-F(2x)\right]F(x)+2(T-T')\int_{x}^{2x}dy\frac{\partial F}{\partial y} F(y-x) \\
	      &+&Q'\left[1+F^{2}(x)-2F(x)F(2x)\right]-2Q\left[1-F(2x)\right]F(x)+2(Q-Q')\int_{2x}^{\infty}dy\frac{\partial F}{\partial y} F(y-x)\nonumber
\eea

\noindent As a check we see that letting $P'=P$ and $T'=T$ we recover the single rule model. This equation has all the complicating
adjectives one can attach, the most important being non-locality, and a direct approach as that of solving for characteristics is not obvious. Nevertheless the dependence on $x/2$ and $2x$ in   (\ref{eq:yuh})  are suggestive and have a rather simple interpreation; an agent with point $x$ is {\bf protected against mergers} of two others if the points of those are both smaller than $x/2$ and similarly an agent with point $2x$ is protected against mergers of two agents with points  less than $x$. 

One can also show that the ansatz $F(x,t)=F(z\equiv x/t)$  of the single rule model is still applicable here\footnote{This is a consequence of the fact that we have $W(ax,ay,ay')=W(x,y,y')$ for either the single rule or the merger model. This in turn means that the equation has $x\to ax$ and $t\to at$ symmetry.}. The only concern could be the integrals but they are easily transformed accordingly. For instance,

\be
\int_{x_{1}}^{x_{2}}dy\frac{\partial F}{\partial y}F(x-y)=\int_{z_{1}}^{z_{2}}d\zeta F'(\zeta)F(z-\zeta)\;,
\ee

\noindent where $\zeta=y/t$. 
 
\begin{figure}[t]
\vspace*{1.5cm}
\includegraphics[scale=0.5]{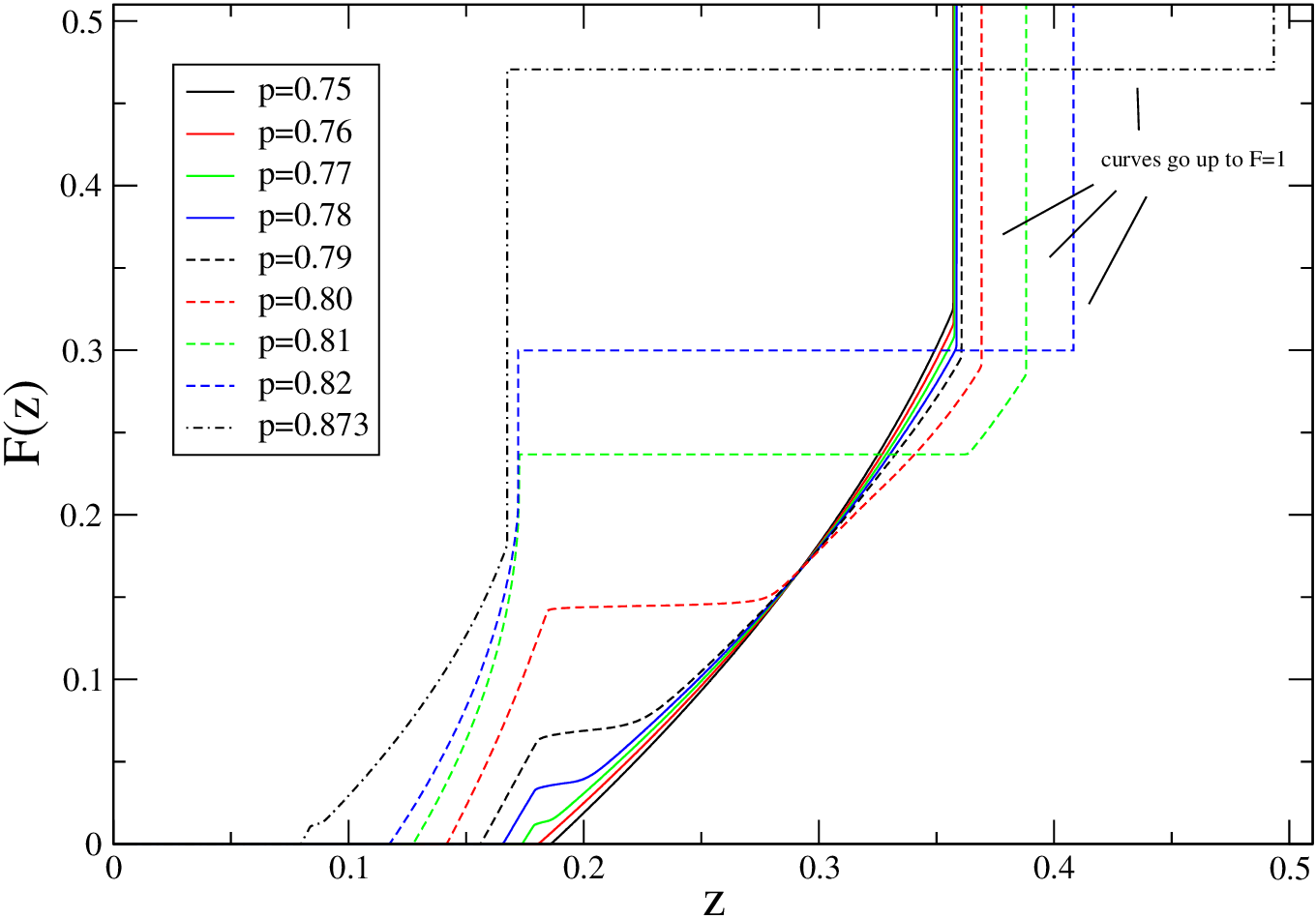}
\caption{\label{fig:1}Results of the numerical simulation for various values of the competitiveness parameter $p$. For a detailed analysis we refer the reader to the text.}
\end{figure}

\subsection{Numerical Analysis}

In view of the obvious difficulty of the equation we resorted to numerical simulation with the hope that it will give us clues about the system. We have performed extensive numerical studies on the system. To be spesific we simulated the microscopic system from $p=0.750$ to $p=1.000$ in steps of $0.001$. For each of these the run was consisted of $N=10^{6}$ agents and we ended the simulation when on average each agent had played about 
$6.5\times 10^{6}$ games meaning, in our normalization, that we stopped the run when time variable $t$ is $2\times 10^{7}$. For all these
cases the initial condition was $F(x,0)=\Theta(x)$.

To start the exposition of numerical results we invite the reader to analyze Fig.\ref{fig:1}. One important aspect one can immediately realize is that upto about $p=0.76$ the resulting distribution is exactly the same as a pure $C^{+}_S$ game. This can only happen if the game is dominated with mergers and no-mergers never happen; so essentially we have a single rule game. Let us call the location of the jump $z_{R}$ and denote point where $F$ vanishes as $z_{L}$. Mergers can {\em gloablly} dominate  if and only if $z_{R}>2 z_{L}$ which simply implies that for any choice of three agents $M+S>L$. So for this case we can use the results of the single rule model to predict when $z_{R}=2z_{L}$ which yields $p_{0}=(\sqrt{13}+1)/6\approx0.76759$ in accordance with the simulations. After this point we see that the number of agents in the vicinity of $z_{L}$  increase \footnote{Needless to say $z_{R}$ and $z_{L}$ are changing as $p$ increases. As can be expected on general grounds $z_{R}$ moves to the right and $z_{L}$ moves to the left. }.
This happens because after the transition the agents accumulated at the jump discontinuity becomes protected againts mergers of the agents around $z_{L}$ and some agents slide down the slope. These agents able to merge no more against the the bunch at $z_{R}$  lag faster than before. Increasing $p$ emphasizes this effect. Later on during the excursion to higher values of $p$, to the right of $z_{L}$  a flat plateau appears. Let us call the position of the left tip of this plateau as $z_{P}$. The emergence of this plateau coincides with the condition $z_{R}=2z_{P}$ at around $p=0.801$. This means that the bunch condensed on the discontinuity at $z_{R}$ becomes completely protected from mergers to the left of $z_{P}$. Furhtermore since $z_{P}>2z_{L}$ the games played among agents to the left of $z_{P}$ are all merger dominated and thus {\em locally}\footnote{That is if we consider the games only between agents in the left bunch.}  of
type $C^{+}_{S}$. Increasing $p$ further we see that another condensation of agents occurs at $z_{P}$, which itself shifts to the left as $p$ takes on higher values. We have turned around full circle and the solution consists of a right bunch condensed at $z_{R}$ and a left bunch sitting at values less than or equal to $z_{P}$. The local structure of this left bunch is similar to the the whole structure
when $p$ was less than $p_{0}$. As the reader could have guessed a further excursion to higher values of $p$ repeats the structure and we end with a
seemingly self similar pattern. Our numerical analysis indicates that as $p$ approaches this pattern repeats itself ad infinitum.

\begin{figure}[t]
\vspace*{1.5cm}
\includegraphics[scale=0.5]{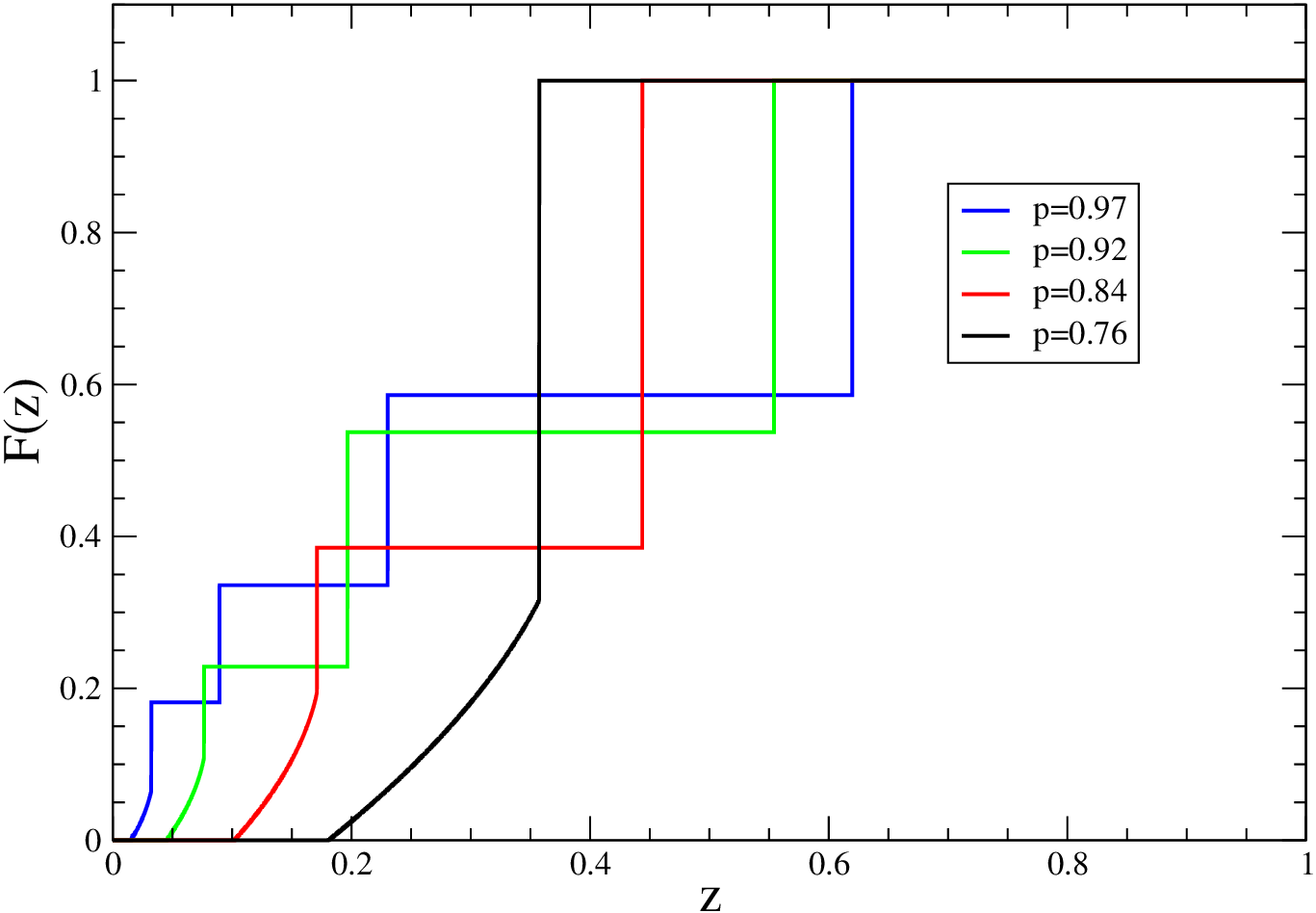}
\caption{\label{fig:2} The qualitatively stable (see text) solutions of the merger model for various number of bunches.}
\end{figure}

To recapitulate we refer to Fig.\ref{fig:2} where we have provided representative solutions relevant to our discussion. There the solutions we have presented
are those that have a qualitative stability, that is the location and the heights of the shocks and the form of the leftmost group will alter as we change $p$ but the number of shocks will be constant for a while during such an excursion. We shall call these shocks {\em bunches} and label them as $B_{n}$ with $n=0,1,2\cdots$ were $n=0$ representing the rightmost bunch. The leftmost bunch is not a pure shock but a combination of a shock and a rarefaction wave since it is locally a $C^{+}_{S}$ solution. The transitions between
those solutions are complicated and bear the full complexity of the differential equation which is somewhat unpenetrable. As can inferred from Fig.~\ref{fig:2}, the most important aspects of these solutions are

\begin{itemize}
 \item The games among players of a single bunch $B_{n}$ are all mergers, since the points of these agents are localized around a particular value.
 \item All players in a bunch $B_{n}$ are protected from mergers of two players in $\bigcup_{k>n}B_{k}$, since the localized points of each bunch is larger than twice that of those to the left of it. 
\end{itemize}

\subsection{Self similar behaviour of the model}

In view of the discussion above, for values of $p$ larger than about $0.81$, we can separate the equation into two parts, one valid near the condensation of the rightmost bunch (the forrunner agents) and the rest. All we need is the interaction of these two parts. From the numerical study we know that the rightmost
bunch is completely protected from mergers of two agents not in this bunch and that the games within this bunch are merger dominated. Let us
start with the generic form of the eqution

\be
\frac{\partial F}{\partial t}=-\frac{\partial F}{\partial x}\int\int dy dy' W(x,y,y')\frac{\partial F}{\partial y}\frac{\partial F}{\partial y'}\;.
\ee

\noindent  As we have mentioned for the merger model we can recast this in terms of $z=x/t$ 

\begin{subequations}
\bea
&&\frac{dF}{dz}\left(-z+\mathcal{G}'[F]\right)=0\\
\mathcal{G}'[F]&=&\int\int dy dy' W(x,y,y')\frac{\partial F}{\partial y}\frac{\partial F}{\partial y'}\;,
\eea
\end{subequations}

\noindent where the integrals are over the whole domain of points.

 Let us denote the rightmost bunch as $B_{0}$ and all the agents not in this bunch as $B_{L}$.
The points of the agents in $B_{0}$ are coalesced  near a value $x_{0}(t)$ which is at least twice as large as the largest point of the agents in $B_{L}$. Therefore the equation for $B_{0}$ can be written as

\be
\frac{\partial F_{0}}{\partial t}=-\frac{\partial F_{0}}{\partial x}\int\int dy dy' W(x,y,y')\frac{\partial F}{\partial y}\frac{\partial F}{\partial y'}
\ee

\noindent where in evaluating $W$ we should remember that $x$ is near $x_{0}$. We end up in

\be
\frac{\partial F_{0}}{\partial t}=-\frac{\partial F_{0}}{\partial x}\mathcal{G}'_{0}
\ee

\noindent with

\bea
\mathcal{G}'_{0}=&&P'(F_{0}-\Phi_{1})^{2}+2T'(F_{0}-\Phi_{1})(1-F_{0})+Q'(1-F_{0})^{2}\\
&&+2P'\Phi_{1}(F_{0}-\Phi_{1})+2T'\Phi_{1}(1-F_{0})\nonumber\\
&&+P\Phi_{1}^{2}\nonumber
\eea

\noindent  The value $\Phi_{1}$ represents the total number of agents in $B_{L}$. That is, $F_{0}$ starts at $\Phi_{1}$ to the left of $B_{0}$ and ends at $1$ to the right. 
The first line above are the self games of the rightmost bunch. The second line
represents games where two agents are in $B_{0}$ and one in $B_{L}$. The third line represents games where only one agent is selected from
$B_{0}$. 

For values of $x<x_{0}/2$ we are in $B_{L}$. Denoting the cumulative function in this region as $F_{L}(x,t)$ we get

\be
\frac{\partial F_{L}}{\partial t}=-\frac{\partial F_{L}}{\partial x}\mathcal{G}'_{L}
\ee
\noindent with 

\bea\label{eq:33}
\mathcal{G}'_{L}= &&\int\int dy dy' W(x,y,y')\frac{\partial F_{L}}{\partial y}\frac{\partial F_{L}}{\partial y'}\\
&&+2T(1-\Phi_{1})F_{L}+2Q(1-\Phi_{1})(\Phi_{1}-F_{L})\nonumber\\
&&+Q'(1-\Phi_{1})^{2}\nonumber\\
\eea

\noindent Now $F_{L}$ starts from $0$ and ends at $\Phi_{1}$ and the second (third) lines in the equation above represents picking two (one) agents from $B_{L}$. The first line represents the self games of all the agents that are in $B_{L}$.

There is resemblance to self-similarity in (\ref{eq:33}); the first line looks like the original equation but the interaction terms with the bunch $B_{0}$ spoils this as $B_{L}$ can still gain points via these terms. However in the extreme competitiveness limit $p\to 1$ these terms
are absent since $T$, $Q$ and $Q'$ all vanish. That is, agents in $B_{L}$ will only gain points againts themselves and will simply
remain idle during any competition with the bunch $B_{0}$. Conversely bunch $B_{0}$ will use $B_{L}$ as a definite source of points. In this 
limit (\ref{eq:33}) becomes

\be
\mathcal{G}'_{L}= \int\int dy dy' W(x,y,y')\frac{\partial F_{L}}{\partial y}\frac{\partial F_{L}}{\partial y'}
\ee

\noindent Let us recall however that the maximum value $F_{L}$ can take is $\Phi_{1}$. Defining $\tilde{F}_{L}\equiv F_{L}/\Phi_{1}$ and using the ansatz $z=x/t$ for a solution the equation becomes

\be
\frac{d\tilde{F}_{L}(z)}{dz}\left[-\frac{z}{\Phi_{1}^{2}}+\mathcal{G}'_{L}\right]=0
\ee

This has exactly the same form as the original equation if we let $z\to \Phi_{1}^{2}z$ which one could interpret as scaling of $x$. However unless the initial condition can be partitioned this way we can not say that the solution will resolve itself into a self-similar shape. The inital data we use $F(x,t)=\Theta(x)$ can be partitioned this way because $\Theta(ax)=\Theta(x)$. In conclusion {\bf if $F(x,0)=\Theta(x)$ we expect self-similar behaviour in the solution when $p=1$ via the scaling we have described above}. The procedure of extracting the rightmost bunch is somewhat similar to renormalization procedure and the decoupling mechanism in field theory where after integrating out high energy degrees of freedom we end up with a new theory. If the original theory is said to be  non-renormalizable the new theory is different. If otherwise the
new theory is similar in form to the original except quantities in it like fields, coupling constants etc. are scaled it is called a renormalizable theory. We see an analogy here; the extraction of the rightmost bunch yields the same form of equations for $p=1$  and different otherwise. Thus within this sense we can say that for $p=1$ and with $F(x,0)=\Theta(x)$ the model is renormalizable. The effect
of this renormalization yields the scaling of $F$ via $F\to \Phi_{1} F$ and that of $z$ via $z\to z\Phi_{1}^{2}$.

Given these circumstances we can repeat the same structure infinitely many times where we end up with infinitely many
bunches $B_{n}$ localized around $z_{n}$ containg $f_{n}\equiv\Phi_{n+1}-\Phi_{n}$ agents. And as expected at each step the scaling should be via the same number $\Phi_{1}\equiv\varphi$ if there is to be self-similarity at all. The protection of $B_{n}$ againts mergers of any two players in $B_{k>n}$ mandates that $z_{n}>2z_{n+1}$. Reiterating this procedure we find the equation obeyed for agents in $B_{n}$ to be

\be
\frac{dF_{n}}{dz}\left[-z-2F_{n}(z)^{2}+2F_{n}(z)\Phi_{n}+\Phi_{n+1}^{2}\right]=0\;.
\ee

\noindent  As expected this can only be resolved via a shock, the location of which is found via the Rankine-Hugoniot condition

\be
z_{n}=\frac{1}{3}(\Phi_{n+1}^{2}+\Phi_{n}\Phi_{n+1}+\Phi_{n}^{2})\;.
\ee

\noindent Since at each step we scale with the same number $\varphi$ we have

\be
\Phi_{n}=\varphi^{n}\;.
\ee

\noindent Which in concert with the protection mechanism mentioned means that $\varphi<1/\sqrt{2}$ and implies the following

\begin{subequations}
\bea
z_{n}&=&z_{o}\varphi^{2n}\;,\\
z_{0}&=&\frac{1}{3}(\varphi^{2}+\varphi+1)\;.
\eea
\end{subequations}

We are one equation away from a solution. A concept we may use is the stability of the solution at large times. For instance let
us focus on the second bunch $B_{1}$. To the right of this bunch there is $B_{0}$ containg $1-\varphi$ agents and to the left of it there
are infinitely many bunches accomodating $\varphi^{2}$ elements. Now if the solution is stable the location of $B_{1}$ in terms of the variable $z$ does not change in time. That is the points they loose against $B_{0}$ is the same as the points they gain from all the agents to the left. This
can only happen if these two regions have the same number of agents yielding $\varphi^{2}=1-\varphi$. One can similaryly argue as follows; how can the bunch $B_{0}$ knows that it is the leading bunch in a self-similar pattern? Let us assume the existance of a further bunch $B_{-1}$. Using the scaling we expect the number of agents in this bunch to be $1/\varphi -1$. Now the location of $B_{0}$ will be stable if and only if the games lost to $B_{-1}$ only by $B_{0}$ equals the games won against agents in $\cup_{k>0}B_{k}$ again only by $B_{0}$. This can only happen if the number
of agents in $B_{-1}$ equals the number of agents below $B_{0}$ meaning $1/\varphi -1=\varphi$. These two considerations are equivalent
and allows us to find

\begin{subequations}
\bea
\varphi=\frac{\sqrt{5}-1}{2}\;,\\
z_{0}=\frac{2}{3}\;.
\eea
\end{subequations}

\noindent Thus $\varphi$ is the reciprocal of the Golden Ratio and $z_{0}$ is just twice the value of the mean
speed of the entire system. So a player in the rightmost bunch is, {\em in the mean}, on the verge of being protected from mergers of {\em any two} 
randomly picked agents from the entire collection. The simulation results for which $F(x,0)=\Theta(x)$ are in very good aggreement with this theoretical prediction.

\begin{figure}[t]
\vspace*{1.5cm}
\includegraphics[scale=0.5]{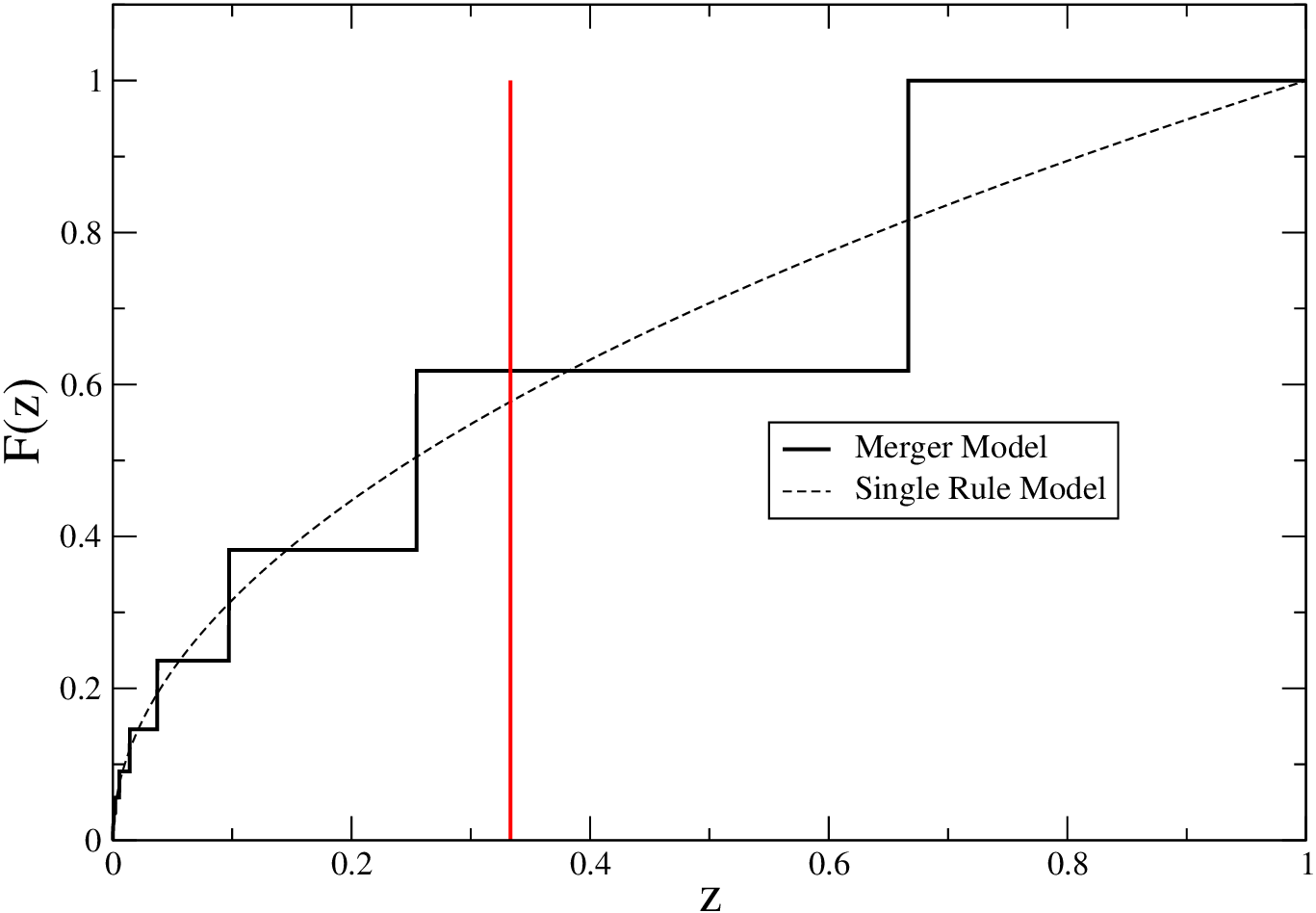}
\caption{\label{fig:4}The solution to the merger model in comparison to that of the single rule model both for $p=1$. The mean speed of
agents is shown with a vertical line.}
\end{figure}

It is interesting to contrast the model with mergers and the single rule model without mergers. The comparison is in Fig.~\ref{fig:4}.
One important aspect we realize is that the effect of implementing mergers does not in general affect the global behaviour of the problem; the model with mergers is like a discretization of the curve $F(z)=\sqrt{z}$ so overall competitiveness is still there. However the local behaviour is completely different; we have {\bf stratification} of agents. It is rather peculiar to observe this behaviour when we, in effect, increased the overall competitiveness of the model. That is, one can say that the environment is now subject to more coercive rules of competition which in contrast leads to condensation of agents around particular values of point gain. As we have mentioned before this effect is a consequence of the conflict between mergers living in the $C^{+}_{S}$ regime and no-mergers being in the $C^{-}$ class of solutions were the simulations run unconditionally as a merger or no-merger single rule model respectively.

\subsection{Restricted Mergers}

The merger model we have presented represents a very coercive environment of competition and it is not readily susceptible to analytical
study for arbitrary $p$. The main reason for this is its high non-locality. This non-locality is present because the two lowest laggers are allowed to merge in all cases even when $M+S=L+1$. One could wish to contemplate other schemes of mergers where this effect is less pronounced. One way to do this is to regulate mergers. Here we present a model which is the most restricted. Let us remember that at time $t$ the theoretical maximum point is just $t$. Now let us pick three agents and order their points as $L\geq M\geq S\;$ and  {\em let us allow mergers only if $S>t/2$}; this makes sure that the competition is a merger unit since we always  have $L<t$ and thus it is always true that $M+S\geq 2L$. In such an approach mergers
will be represented in $W$ via a term like $\Theta(x-t/2)$ wich will become $\Theta(z-1/2)$ for the asymptotic behaviour where as usual $z=x/t$. The equation becomes

\be
\frac{dF}{dz}\left[-z+G'(F)\right]=0\;,
\ee

\noindent with

\be
G'(F)=\left\{
 \begin{matrix}
PF^{2}+2TF(1-F)+Q(1-F)^{2} \;\;\;&&{\rm for}\;\;\; z<1/2\\
PF^{2}+(P'-P)(F-\bar{F})^{2}+2T\bar{F}(1-F)+2T'(1-\bar{F})(F-\bar{F})+Q'(1-\bar{F})^{2}\;\;\;&&{\rm for}\;\;\;z\geq 1/2
\end{matrix} \right. 
\ee

\noindent and with $\bar{F}\equiv F(1/2)$. This equation is local and thus can be studied analytically in much the same way as the single rule model. Here we only present the solution for $p=1$ to compare it with the unrestricted merger model.

\be
F(z)=\left\{ \begin{matrix}
            \sqrt{z}										&& z\leq \frac{1}{2} \\
            \frac{1}{2}\left(\sqrt{2}+1-2\sqrt{1-z}\right)  && \frac{1}{2} \leq z \leq z_{r} \\
            1												&& z_{r} \leq z \\
			\end{matrix}
			\right.
\ee

\noindent where we have 

\bea
\bar{F}&=&\frac{1}{\sqrt{2}}\;,\nonumber\\
z_{r}&=&\frac{2\sqrt{2}-1}{2}\;.\nonumber
\eea

\begin{figure}[t]
\vspace*{1.5cm}
\includegraphics[scale=0.5]{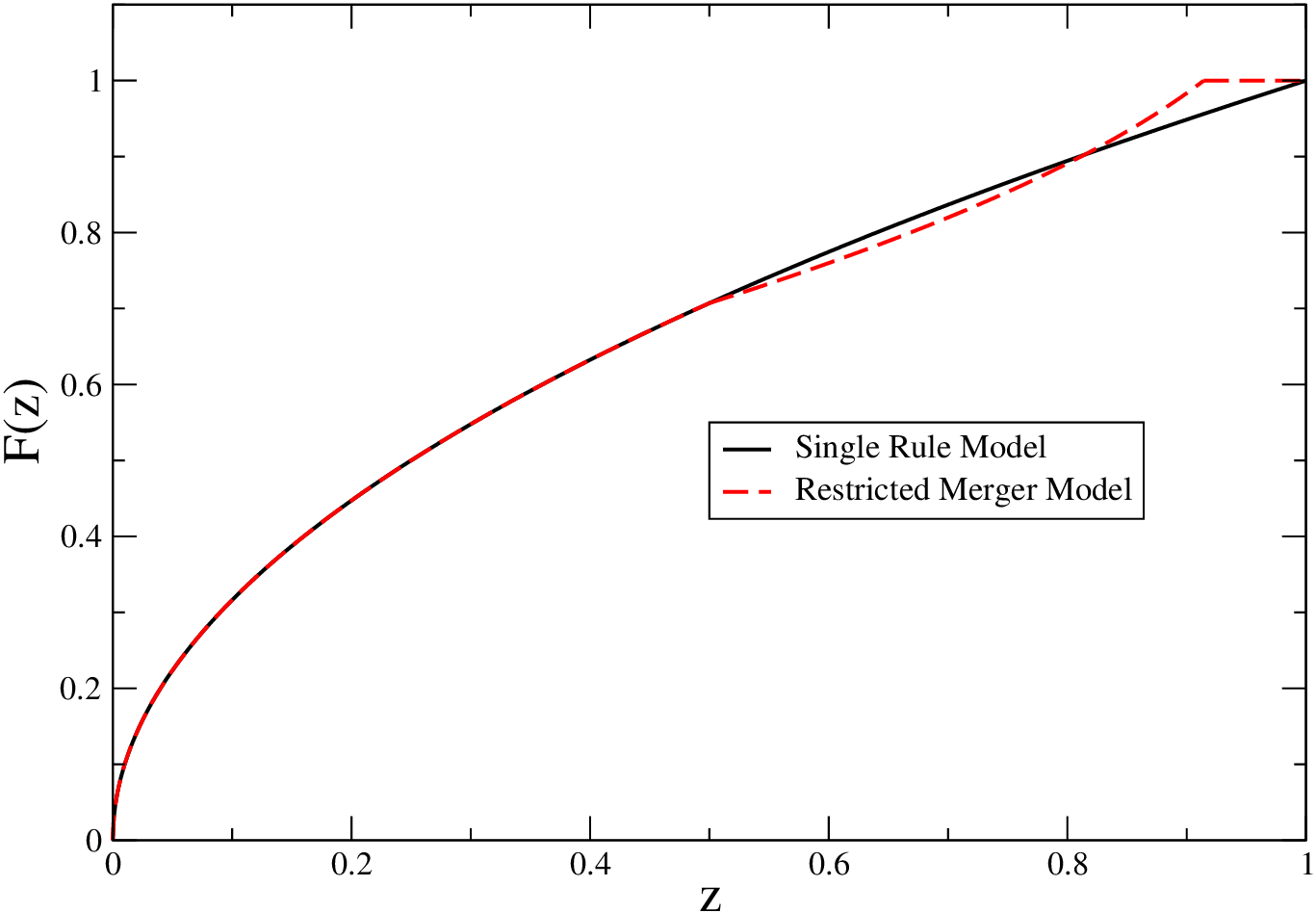}
\caption{\label{fig:5}The solution to the restricted model case in comparison with the curve for single rule model both for $p=1$}
\end{figure}

\noindent Which is in very good agreement with numerical simulations. As we see the solution for $z<1/2$ is the same as that of the single
rule model. The comparison for the full range of $z$ is given Fig.~\ref{fig:5}. As the reader could have guessed the stratification effect
is non-existent but the tendency of the curve to go to that regime had mergers were unrestricted is apparent.

\section{Digression on Initial Conditions}

Let us remember that the generic form of the equation governing the dynamics of three agents games is

\bea
\frac{\partial F}{\partial t}&=&-\frac{\partial F}{\partial x}G'[F]\nonumber\\
G'[F]&\equiv &\int\int\;dy\;dy'\;\frac{\partial F}{\partial y}\frac{\partial F}{\partial y'}W(x,y,y')\;,\nonumber
\eea

For the single rule model of Chapter I. the integrals resolve into a simple polynomial of $F$. The reason for such a simplification is, for the model mentioned, the fact that $W$ being only a function of the ordering of points containing only terms like $\Theta(L-M)\Theta(M-S)$ and thus obeying 

\begin{subequations}
\bea
W(ax,ay,ay')&=&W(x,y,y')\label{eq:sym1}\\
W(x-b,y-b,y'-b)&=&W(x,y,y)\label{eq:sym2}
\eea
\end{subequations}

\noindent The first of these equations means that the equation will be invariant under the combined transformations $x\to ax$ and $t\to at$ which makes it possible to assume an ansatz $F(x/t)$ since $F(x,0)=\Theta(x)$ also obeys this symmetry. 

Now let us shift $F$ by a constant $\phi$ such that $F=\tilde{F}+\phi$ with $\phi=-(T-Q)/(1-3T)$. Since the single rule model obeys (\ref{eq:sym2}) one can also perform a Gallilean transformation\footnote{Under the Gallilean transformatio alone the equation is transformed into $\frac{dF}{dz}\left[-z+z_{0}+G'(F)\right]$. That is the $z$ variable which in principle represents the speed of agents
is shifted by a constant.} on the independent variables of the form $\tilde{t}=t$ and $x=\tilde{x}-z_{0}\tilde{t}$. Choosing $r=Q-(T-Q)^{2}/(1-3T)$ the equation becomes

\be
\frac{\partial \tilde{F}}{\partial \tilde{t}}=-(1-3T)\tilde{F}^{2}\frac{\partial \tilde{F}}{\partial \tilde{x}}\nonumber\;,
\ee

\noindent which can be recast as

\begin{subequations}\label{eq:eben}
\bea
\frac{\partial \tilde{F}}{\partial t}&=&-\frac{\partial \tilde{G}(\tilde{F})}{\partial x}\;,\\
\tilde{G}(\tilde{F})&=&\frac{(1-3T)}{3}\tilde{F}^{3}\;.
\eea
\end{subequations}

\noindent where $\tilde{G}(\tilde{F})$ is strictly concave for $(1-3t)>0$. If on the other hand $(1-3t)<0$ one can make $\tilde{G}(\tilde{F})$ strictly concave by doing $\tilde{x}\to -\tilde{x}$.

After these transformations the initial condition is changed into $\tilde{F}(\tilde{x},0)=\Theta(\tilde{x})+\phi$ which still represents 
a Riemann problem. Now it is a known fact that the solutions to equations of type (\ref{eq:eben}) converge in the infinite time limit to the solutions of the Riemann problem\footnote{See for instance \cite{Lax}, \cite{Matsu1} and \cite{Nishi1}} if the initial condition obeys $\tilde{F}(\tilde{x},0)=F_{L}$ for some $x<x_{L}$ and $\tilde{F}(\tilde{x},0)=F_{R}$ for some $x>x_{R}$ where $F_{L}$ and $F_{R}$ are constants. Now the dependent variable $F$ is the
cumulative of a globally conserved quantity; the number of agents. Therefor for a generic initial distribution of points we have $F(x,0)=0$ for $x<0$ and $F(x,0)=1$ for some $x>x_{R}$.  We thus infer that {\bf for the single rule model the time asymptotics of $F$  is independent of the initial conditions}\footnote{For generic initial data $F(x,0)$ of the type mentioned in the text the simulations converge to that of $F(x,0)=\Theta(x)$ after a comparatively larger number of Monte-Carlo cycles. An estimate of this time is presented for instance in \cite{naka1}.}. These considerations also apply to the restricted merger model we have studied since in principle it has the same general form as the single rule model.

For the merger model without restrictions we still have the symmetry in (\ref{eq:sym1}) which allows us to make the $F(x/t)$ ansatz if the initial condition obeys $F(ax,0)=F(x,0)$. Unfortunately the shift symmetry in (\ref{eq:sym2}) is absent because mergers are implemented via terms of the type $\Theta(L-M)\Theta(M-S)\Theta(M+S-L)$. This makes the equation highly non-local and in particular the 
Gallilean transformations will take it to an entirely different form. None of the theorems presented in the mentioned papers above hold
and one would expect a strong dependence of the time asymptotics on the initial data; an observation which we have substantiated with
numerical simulations.

\section{Discussion}

The unrestricted merger model we have presented has interesting properties. The most important being the stratification of the entire society of agents. The bunch that has the largest number of agents (this number is $(1-\varphi)\approx \% 32$) 
is also the bunch with the largest rate of point gain. However this point gain is only two thirds of the maximum possible rate. On the
other hand all the agents other than the first bunch are earning slower than the mean rate. The number of agents living below this mean is
slightly lower in the single rule model where mergers are not implemented. Furthermore from our solution it is clear that the number of
agents in a bunch $f_{n}$ satisfies $f_{n}=f_{n-1}-f_{n+1}$, that is the number of agents in a bunch is like a derivative in the sense
of the bunches. 

It is tempting to speculate that the merger model we have presented could have applications to natural or social phenomena. The stratification phenomenon being present in various systems. For instance one could argue that a bunch, in essence, represents a single entity, the number of agents in it simply meaning that it has more activity in taking part in games. From this perspective one may interpret
the merger model as one of explaining monopoly formation after a period of competition between companies. 
Stratification is also present in natural systems. Another tempting interpretation could be
the stratification of the collection of entire living species in terms of their genetic material. If there is a competition mechanism
complexified with mergers like the one described in this work one could hope to gain qualitative understanding of the formation of different strata of living
organisms. Agents could be units of genetic material  and the competition could be for taking part in the genetics of a (new) species. 

The emergence of the stratification mechanism can be interpreted in the following way. The rules of mergers yield a complicated and non-local system. The equations are so complicated that they can not be resolved via smooth functions and the only possible escape is the formation of various shocks; there {\bf must} be a solution since we are simply simulating a Monte-Carlo system with a well defined microscopic model.  We believe this to be true for other systems of conservation laws, coming from well defined microscopics, where the equations become non-local.

\end{document}